\pdfoutput=1
\documentclass[aip,amsmath,amssymb,reprint]{revtex4-1}

\usepackage{graphicx}
\usepackage{dcolumn}
\usepackage{bm}
\usepackage{wrapfig}
\usepackage{tikz}
\usepackage{graphicx}
\usepackage{subfigure}
\usepackage{blindtext}
\usepackage{textcomp}
\usepackage{makecell,multirow}
\usepackage{amsmath}
\usepackage{amsfonts}

\begin{document}
\preprint{AIP/123-QED}
\title{In-Plane Plasmon Coupling in Topological Insulator Bi$_2$Se$_3$ Thin Films}

\author{Saadia Nasir$^1$, Zhengtianye Wang$^2$, Sivakumar V. Mambakkam$^2$, and Stephanie Law }
 \homepage{slaw@udel.edu}
\affiliation{Department of Physics and Astronomy, University of Delaware, 217 Sharp Lab, 204 The Green, Newark DE 19716 USA}
\affiliation{Department of Materials Science and Engineering, University of Delaware, 201 DuPont Hall, 127 The Green, Newark DE 19716 USA}
\date{\today}

\begin{abstract}
The surface states of the 3D topological insulator (TI), Bi$_2$Se$_3$, are known to host two-dimensional Dirac plasmon polaritons (DPPs) in the terahertz spectral range. In TI thin films, the DPPs excited on the top and bottom surfaces couple, leading to an acoustic and an optical plasmon mode. Vertical coupling in these materials is therefore reasonable well-understood, but in-plane coupling among localized TI DPPs has yet to be investigated. In this paper, we demonstrate in-plane DPP coupling in TI stripe arrays and show that they exhibit dipole-dipole type coupling. The coupling becomes negligible with the lattice constant is greater than approximately 2.8 times the stripe width, which is comparable to results for in-plane coupling of localized plasmons excited on metallic nanoparticles or graphene plasmon polaritons. This understanding could be leveraged for the creation of TI-based metasurfaces.

\end{abstract}

\maketitle

Plasmons are the fundamental excitations of free electrons in a system. Localized surface plasmons on the metallic nanoparticles (MNPs) have been investigated by many researchers in the past few decades for their interesting optical properties. These optical properties depends on the particle shape, size, and material. In addition, changing the spacing between the MNPs can tune their optical properties due to dipole-dipole type coupling between MNPs in close proximity. The resonance frequency of excited plasmons has been demonstrated to decay exponentially with decreasing separation length between the nanoparticles\cite{kheirandish2020analytical,jain2007universal}. It has also been reported that when light is shined on metallic nanoparticles with polarization parallel to their long axis, a red shift in the resonance wavelength is observed with increase in gap size between them \cite{rechberger2003optical,atay2004strongly,tabor2009use}. The shift in the resonance frequency is found to be negligible when the separation between MNPs exceeds approximately 2.5 times their short axis length \cite{su2003interparticle}.

In addition to three-dimensional MNPs, plasmons have also been excited in two-dimensional materials including graphene and the surface states of topological insulators (TIs)\cite{stauber2014plasmonics,politano2015interplay,politano2017cutting,kogar2015surface}. TIs are materials that exhibit a bulk band gap that is crossed by surface states with linear dispersion. These surface states exist at the interface between a TI and a trivial material and are occupied by massless Dirac electrons. When light is shined on a TI, two-dimensional Dirac plasmon polaritons (DPPs) can be excited. \cite{di2013observation,stauber2014plasmonics}. The resonance frequency of the DPPs excited in TIs lies in the terahertz (THz) spectral range\cite{zhang2010topological,autore2017terahertz}. Therefore DPPs in TIs have practical applications in THz sensors, waveguides and spintronics \cite{politano2017optoelectronic,tang2018ultrasensitive,viti2016plasma}. Because the wavelength of light with THz frequency is greater than the thickness of TI films, DPPs are excited on both top and bottom surfaces of film simultaneously. They couple to each other through their evanescent electric fields resulting in an optical and an acoustic mode. Only the optical mode can be excited by light, and its dispersion relation can be given by Eq. \ref{DR}\cite{stauber2013spin}.
\begin{equation}\label{DR}
{\omega_p}^2=\dfrac{e^2}{\epsilon_0} \dfrac{v_f\sqrt{2\pi n_D}}{h} \dfrac{q}{\epsilon_T+\epsilon_B+qd\epsilon_{TI}}
\end{equation}

\noindent where $\omega_p$ is the optical DPP frequency, $e$ is the charge of the electron, $v_f$ is the Fermi velocity, $n_D$ is the Dirac charge density, and $d$ is the thickness of film. $\epsilon_0$, $\epsilon_T$, $\epsilon_B$, $\epsilon_{TI}$ correspond to the permittivity of free space, the superstrate, the substrate, and inside the TI thin film, respectively, and $q$ is the wavevector of the light. 

To date, localized TI DPPs have successfully been excited in both stripe and micro-ring geometries, and propagating TI plasmons have been excited in thin films \cite{di2013observation,autore2015plasmon,venuthurumilli2019near,wang2020propagating}. However, in-plane coupling between the localized TI DPPs has yet to be explored. In this paper, we investigate in-plane coupling between localized TI DPPs excited in a stripe geometry. We grew eight 50nm thick films of Bi$_2$Se$_3$, a prototypical 3D TI using molecular beam epitaxy (MBE) and etched them into stripes of width 2.5$\mu m$ with varying edge-to-edge spacing. We then measure the extinction spectra of the samples as a function of stripe width and show clear evidence of in-plane coupling among the DPPs. Understanding this coupling is important both for expanding our understanding of the unusual DPPs in TI films and for future devices that could leverage this coupling such as the creation of metasurfaces.

 
The two-step growth method was used to grow 50nm thick Bi$_2$Se$_3$ thin films on sapphire (0001) substrates in a Veeco GENxplor MBE system. First, the substrates were outgassed in the load lock at 200\textcelsius\, for 10 hours after which they were transferred to the main chamber and heated to 650\textcelsius\,. They were held at that temperature for 5 minutes to desorb any residual impurities. The substrates were then cooled to the growth temperature of 325\textcelsius\,. 5nm of Bi$_2$Se$_3$ was grown using grow anneal strategy where we grew Bi$_2$Se$_3$ for a minute with co-deposition of bismuth and selenium and annealed for one minute under selenium flux for a total of six loops. After the deposition of the seed layer, the substrate temperature was increased to 425\textcelsius\, and the seed layer was annealed for 5 minutes under a continual selenium flux. We grew the remaining 45nm of Bi$_2$Se$_3$ at this temperature before cooling the samples to 200\textcelsius\, under a selenium flux before removing them. Throughout the growth, a 20\textcelsius\,/minute heating and cooling rate was used. Bismuth was supplied from a dual-filament effusion cell where the bulk cell temperature was set to 490\textcelsius\, to obtain a growth rate of 0.74nm/min. The selenium was supplied from a valved cracker source with a cracking zone set to 900\textcelsius\, to improve the incorporation of selenium into the film \cite{Ginley2016}. The Se:Bi flux ratio was ~90-100 as measured by a beam flux monitor. The thickness of the films was confirmed at 50$\pm$1 nm using x-ray reflectivity and the film was confirmed to be in the (0001) orientation by x-ray diffraction measurements (available in the Supplementary Information). Room temperature Hall effect measurements gave a sheet density of (-2.46$\pm$0.01)x10$^{13} $cm$^{-2}$ and a mobility of (-718.48$\pm$2.57) cm$^2$/Vs.\cite{daniel}

\begin{figure}[h] 
	\begin{minipage}[t]{\linewidth}
		\centering
		\subfigure{\includegraphics[scale=0.33]{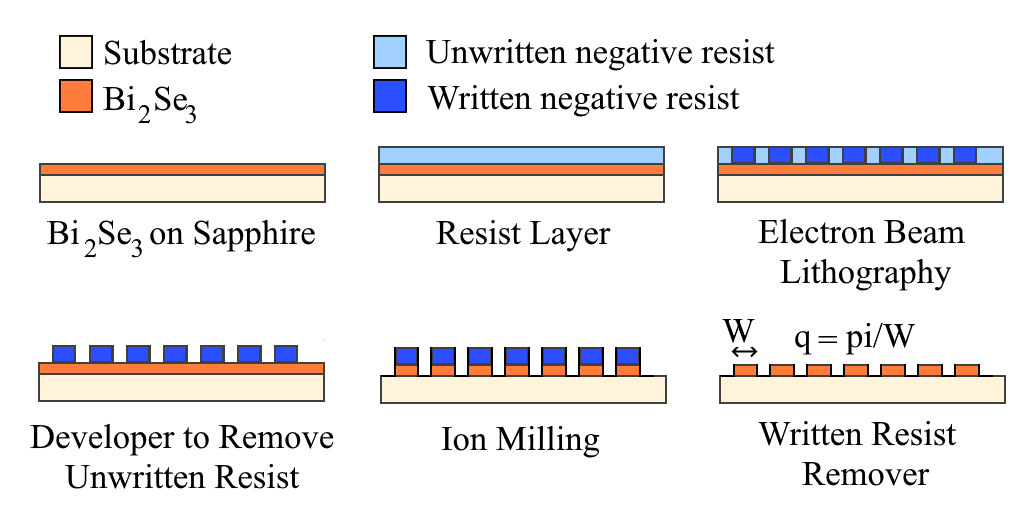}}
	\end{minipage}
	\caption{Steps involved in the fabrication of striped arrays having width W on $Bi_2Se_3$ thin films using EBL. The patterned stripes adds the momentum $q=\dfrac{\pi}{W}$ to the incoming photons.}
	\label{schematic}
\end{figure}

 The films were then patterned into grating structures using electron beam lithography. The stripe width was kept constant for all samples at 2.5$\mu$m while the gap size was changed (0.3, 0.5, 0.7, 1.0, 1.5, 2.0, 2.5 and 7.5$\mu$m). The steps involved in the fabrication of the structures are shown in Fig. \ref{schematic}. We spin the negative resist ARN 7520 on our films to get a coating that is 90nm thick. The film is then exposed in the electron beam lithography system, the resist developed, and the pattern transferred to the film using ion milling. We used NMP/AR 300-76 as solvents to remove the remaining resist, but some resist residue may remain. Room temperature Fourier transform infrared spectroscopy was performed to obtain the extinction spectra of the samples.

\begin{figure}[h]
	\begin{minipage}[t]{\linewidth}
		\centering
		\subfigure{\includegraphics[scale=0.18]{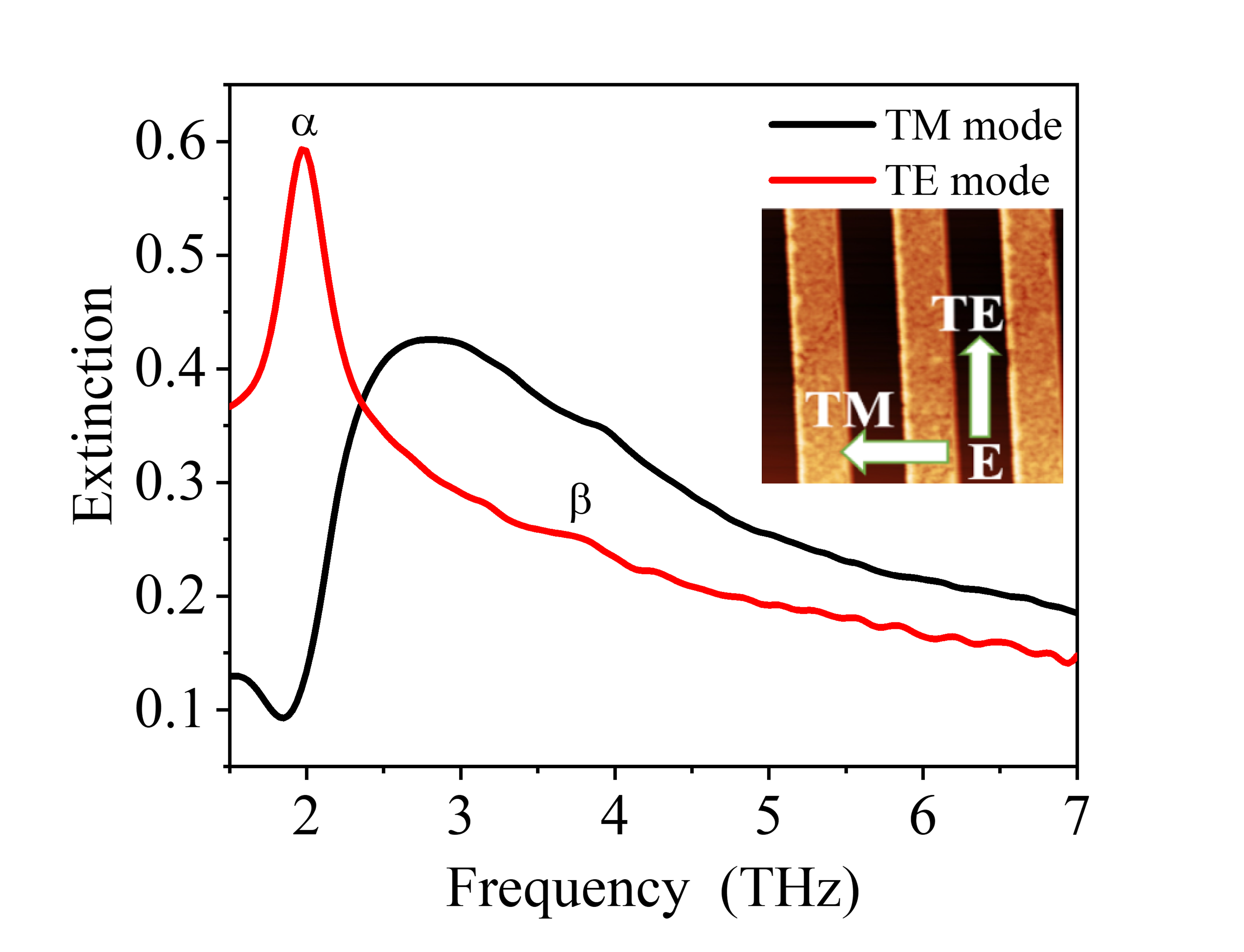}}
	\end{minipage}
	\caption{TE-polarized (red) and TM-polarized (black) extinction spectra of a sample with gap size of 0.3 $\mu m$. Two phonon modes $\alpha$ and $\beta$ are observed in the TE spectrum near 2 THz and 3.9 THz respectively. The inset shows the difference of TE and TM mode. For TE mode, electric field of incident light is parallel to the stripes and for TM mode electric field of incoming photons is directed perpendicular to the film stripes.}
	\label{TETM}
\end{figure}

When performing extinction measurements, light can be incident on samples with its electric field parallel (transverse electric, TE) or perpendicular (transverse magnetic, TM) to the stripes as shown in the inset of Fig. \ref{TETM}. For TE-polarized measurements, the light is unable to excite the plasmons due to the momentum mismatch. For TM-polarized measurements, the photons are able to excite the plasmon mode ,resulting in a standing wave type resonance. In this case, the momentum of the incoming photon is increased by $\dfrac{\pi}{W}$ and plasmon polaritons can be excited \cite{maier2007plasmonics}. Example transverse magnetic (TM) and transverse electric (TE) extinction spectra for the sample with a gap size of 0.3$\mu$m are shown in Fig. \ref{TETM}. These spectra are calculated using 
\begin{equation}\label{extinction}
Extinction = 1-T(\omega)/T_0(\omega) = |e(\omega)|^2
\end{equation}
where $\omega$ is the frequency of the light, $T(\omega)$ is the frequency-dependent transmission through the sample, and $T_0(\omega)$ is the frequency-dependent transmission through the sapphire substrate. In the TE-polarized spectrum (shown in red) we observe two phonon modes: the $\alpha$-phonon mode near 2THz and the $\beta$-phonon mode near 3.9THz. The TE-polarized extinction for all other stripe arrays looks similar and are shown in the Supplementary Information. However, in the TM-polarized spectrum, we see the disappearance of the $\alpha$-phonon mode and the appearance of a broad peak centered near 3THz. This asymmetric lineshape is caused by the excitation of a plasmon polariton mode which couples to the $\alpha$- and $\beta$-phonon modes through a double Fano-type resonance \cite{ginley2018coupled,giannini2011plasmonic}.

\begin{figure}[h]
	\begin{minipage}[t]{\linewidth}
		\centering
		\subfigure{\includegraphics[scale=0.24]{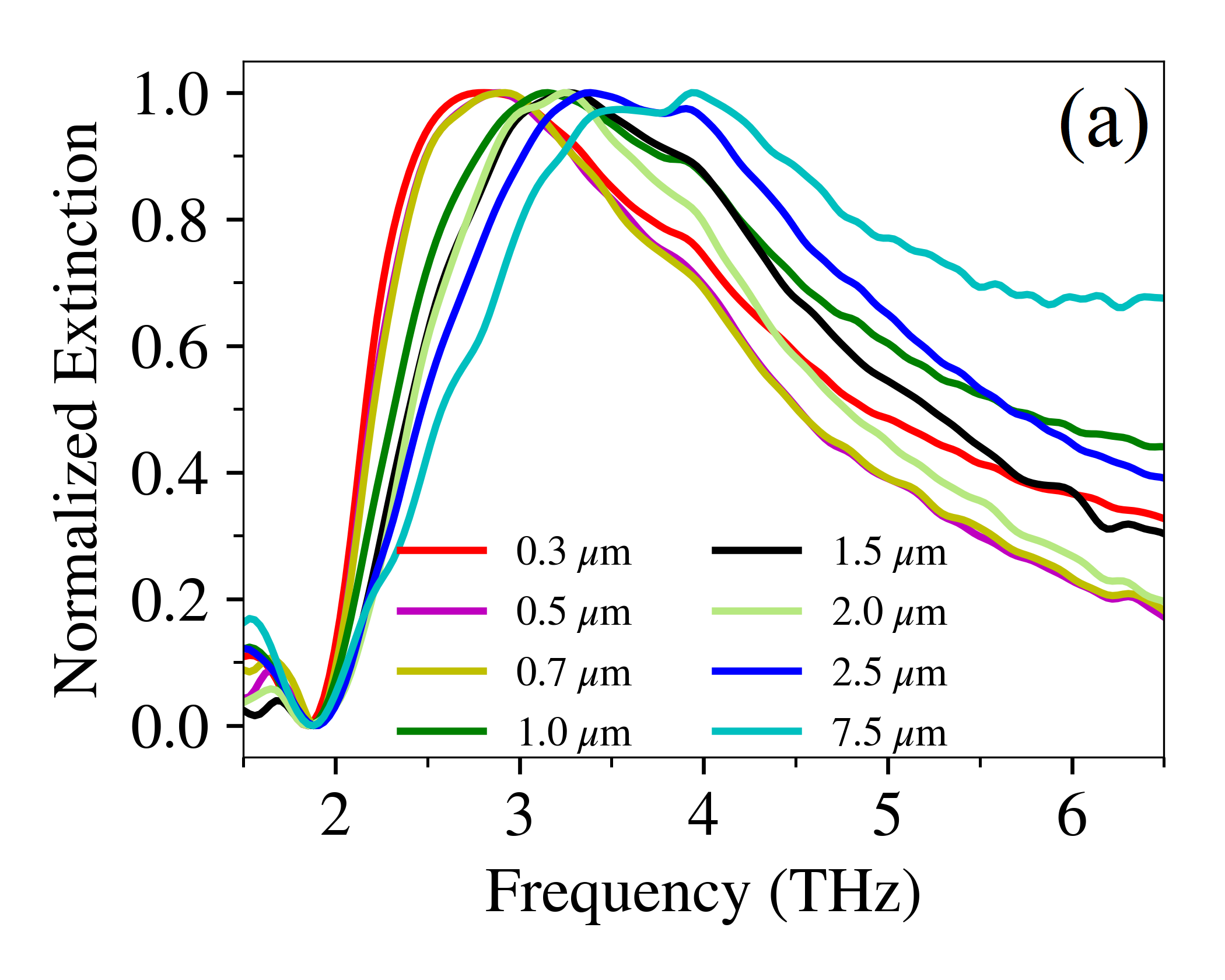}}
		\subfigure{\includegraphics[scale=0.24]{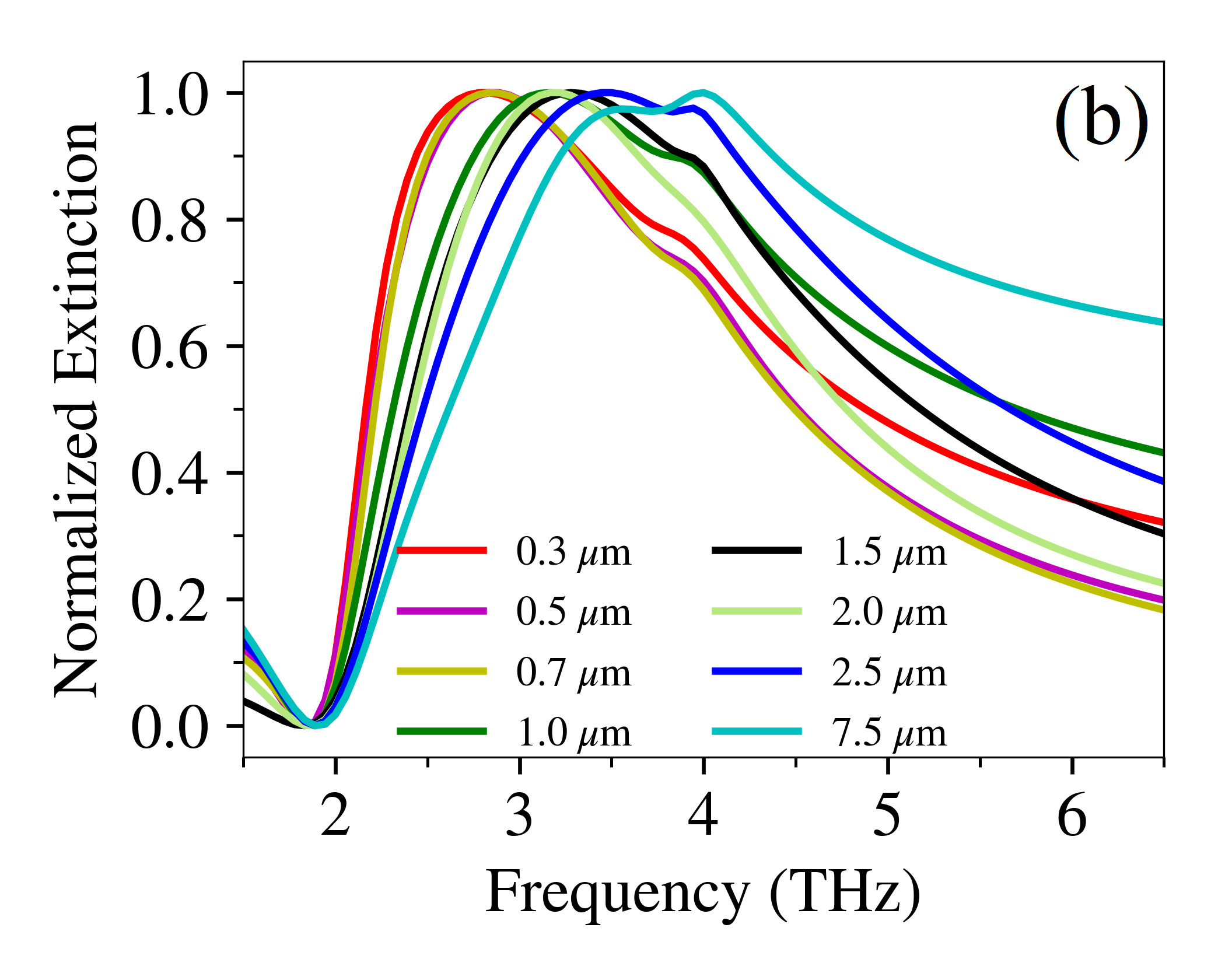}}
		
	\end{minipage}
	\caption{(a) Experimental (b) Fitted normalized TM extinction spectra of all samples. The spectrum shifts towards the higher frequencies with increasing gap size.}
	\label{allsamples}
\end{figure}

The normalized TM-polarized extinction spectra for all eight samples are shown in Fig. \ref{allsamples}(a). We see that the broad plasmon peak is shifting toward higher frequencies as the gap size increases. However, because the plasmon mode is strongly interacting with both the $\alpha$ and $\beta$ phonons, we cannot just pick out the center of the peak as the plasmon frequency. We must instead model the data using a double Fano-type interaction model, as shown in Eq. \ref{Fano} \cite{hao2009tunability,wang2020plasmon}.  
\begin{equation} \label{Fano}
e(\omega)= a_r-\dfrac{\Sigma b_j\varGamma_je^{\iota\phi_ j}}{\omega-\omega_j+\iota\varGamma_j}
\end{equation}
Here, $a_r$ is a constant background term, and for $j^{th}$ oscillator ($j$ is the $\alpha$ phonon, $\beta$ phonon, and plasmon) $b_j$ is the oscillator amplitude, $\phi_j$ is the oscillator phase, $\omega_j$ is the resonance frequency, and $\varGamma_j$ is the line width. We used the non-linear fitting function in Wolfram Mathematica to fit the TM extinction spectra. Each parameter in Eq. \ref{Fano} is left as free parameter within identical defined ranges for all samples shown in black dots in Fig. \ref{frequencyvsgap} (details are in the Supplementary Information). For the sample with a gap size of 0.5$\mu m$, we restricted the $\beta$-phonon linewidth to less than 0.3 THz, and the $\alpha$ and $\beta$ phonon frequencies were fixed to 2.0 THz and 3.9 THz to obtain a better fit. This shifted the value of the plasmon frequency from 2.85 THz to 2.71 THz. For the sample with a gap size of 7.5$\mu m$, the linewidth of the plasmon was restricted to greater than 0.9 THz which pushed the $\alpha$ and $\beta$ phonon frequencies to their edge values of 1.99 THz and 3.91 THz, respectively, but did not effect the plasmon frequency.  The R-squared value for all fittings were greater than 0.999. The modeled extinction curves for all eight samples are shown in Fig. \ref{allsamples}(b). 

\begin{figure}[h]
	\begin{minipage}[t]{\linewidth}
		\centering
	\subfigure{\includegraphics[trim={0.7cm 0.5cm 0cm 1cm},clip,scale=0.37]{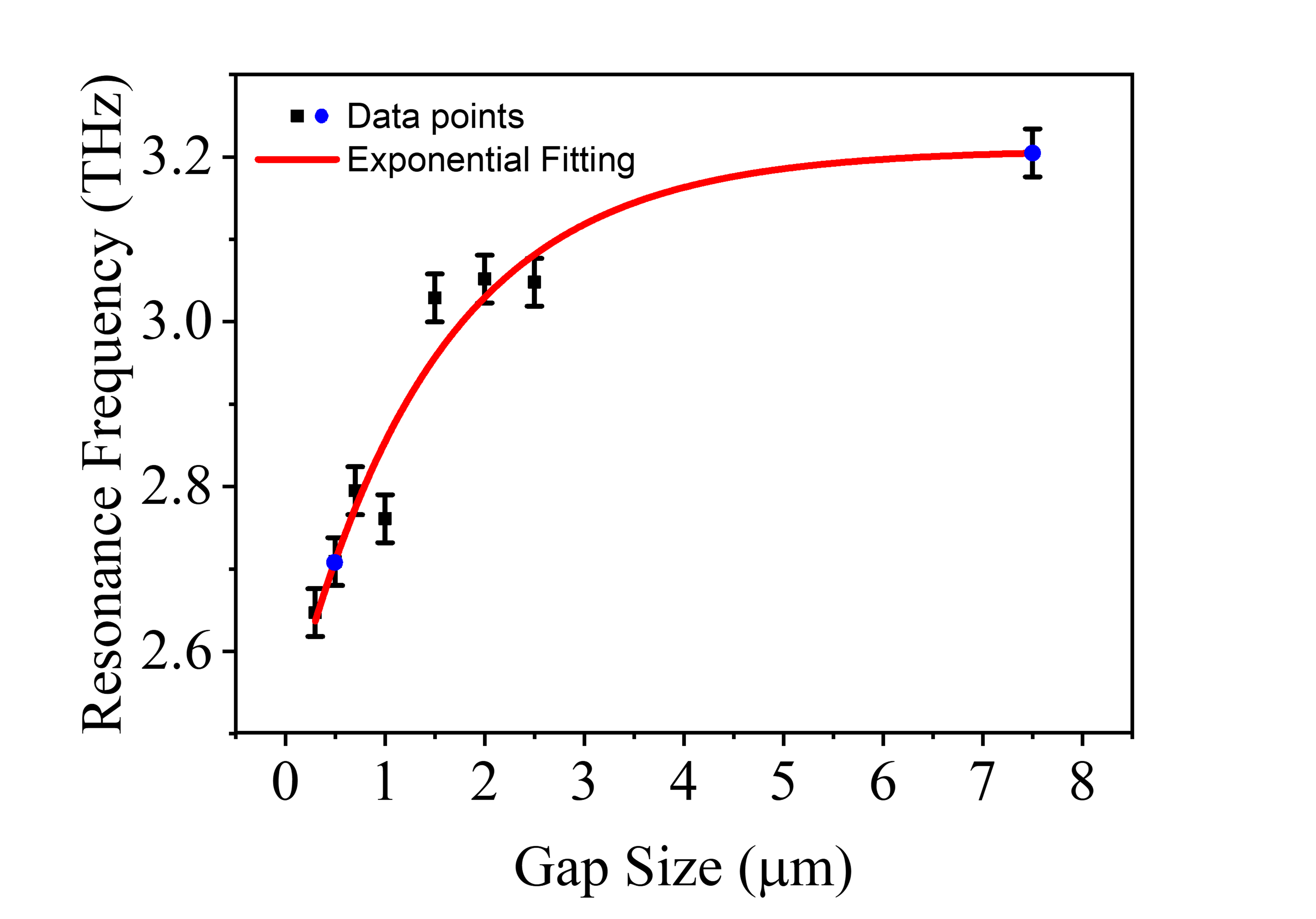}}
	\end{minipage}
	\caption{Extracted plasmon frequencies from double Fano fitting are plotted against respective gap sizes as black squares and blue dots ; these data points are fitted by exponential decay red curve. Three out of thirteen fitting parameters were restricted for data points represented by blue dots. Error bars on the extracted frequencies corresponds to the step size of frequency used for the transmission measurements.}
	\label{frequencyvsgap}
\end{figure}

Finally, in Fig. \ref{frequencyvsgap}, we plot the plasmon frequency extracted from the double Fano fit as a function of gap size. The experimental data is fitted using an exponential decay curve

\begin{equation} \label{exp}
\omega_p = \omega_{pi}- b e^{-{g}/{c}}
\end{equation}

\noindent where g indicates the gap size. The best fit was obtained with a decay constant c= 1.465 $\mu m$ and a saturation frequency $\omega_{pi}$ of 3.208 THz. The parameter b has the units of frequency and its value is 0.702 THz. This parameter along with $\omega_{pi}$ determines the plasma frequency at zero gap size. The increase in the DPP resonance frequency with gap size is a clear indication that we are observing in-plane coupling. As the gap size increases, the interaction between the stripes becomes weaker, resulting in the increase of the resonance frequency. After a certain gap size, the plasmons in the adjacent stripes will no longer couple to each other, and the resonance frequency will saturate at $\omega_{pi}$. For these samples, the plasmon coupling between the stripes becomes negligible for gap sizes larger than $\approx$ 4.52$\mu m$ (for 99$\%$ of $\omega_{pi}$).

DPP coupling in TI stripes can be understood by comparison with the coupled MNPs described above. The precise details of the coupling are likely to be different, since in this case, we are dealing with coupled DPPs in TIs. However, the fundamental physics of dipole-dipole coupling remains the same. The coupling between stripes weakens the restoring force experienced by the electrons in the stripe, causing the resonant frequency to redshift as the gap size decreases and the coupling increases. We observe an exponential dependence on separation, similar to the case of MNPs and the case of coupled graphene nanodisks \cite{Yan2012}. The distance at which the coupling becomes negligible (4.52 $\mu$m) corresponds to a lattice constant of 7$\mu$m, 2.8 times the stripe width. This compares well with the value obtained for the case of MNPs and graphene nanodisks of a lattice constant 2.5 times the particle size. The difference may lie in the way the cutoff was determined or may arise from complications in our system since we have coupling in both the in-plane and out-of-plane directions due to the top and bottom surfaces of the TI thin film.

Because plasmons are excited in our films as a standing waves, we can also think of the in-plane plasmon interaction by analogy with coupling among multiple quantum wells (QWs). If we consider just two coupled QWs, the electron wavefunction of the ground state is symmetric while the first excited state is antisymmetric. The energy of symmetric ground state decreases exponentially as the barrier width decrease and is given as \cite{exponential, harrisonquantum}
\begin{equation} 
E_s = E_0 - \dfrac{8}{2\pi} E_0 \sqrt{\dfrac{E_0}{V_0}} e^{-2a\sqrt{\dfrac{2m V_0}{ \hbar^2 }}}
\end{equation}
where E$_0$ is the ground state energy of uncoupled infinite well, m is the mass of particle,  V$_0$ is the potential between the wells and 2a is the barrier width between the wells. This functional form is similar to the change in the plasmon frequency shown in Eq, \ref{exp}, and the concept of electron coupling among multiple quantum wells can be a useful analogy to coupling among multiple localized DPP.

Overall, these data demonstrate clear evidence for in-plane coupling among DPPs in stripes made from the prototypical TI, Bi$_2$Se$_3$. This lays the groundwork for the creation of devices that leverage this effect such as THz metasurfaces using these structures.

\section*{Supplementary Material}
See \textcolor{blue}{supplementary material} for the x-ray diffraction scan of an unpatterned Bi$_2$Se$_3$ film, atomic force microscopy images of a subset of the patterned Bi$_2$Se$_3$ films, transverse-electric polarized spectra for all samples, and tables of fitting parameters and parameter ranges from Eq. \ref{Fano} that we used for the fitted spectral lines in Fig. \ref{extinction}b.


\begin{acknowledgments}
S. N., Z. W, and S. L acknowledge funding from the U.S. Department of Energy, Office of Science, Office of Basic Energy Sciences, under Award No. DE-SC0017801. S. V. M. and S. L. acknowledge funding from the National Science Foundation, Division of Materials Research under Award No. 1838504. The authors acknowledge the use of the Materials Growth Facility (MGF) at the University of Delaware, which is partially supported by the National Science Foundation Major Research Instrumentation under Grant No. 1828141 and UD-CHARM, a National Science Foundation MRSEC under Award No. DMR-2011824.
\end{acknowledgments}

\section*{Data Availibility}
The data that support the findings of this study are available from the corresponding author upon reasonable request.



\bibliography{bibliography}

\providecommand{\noopsort}[1]{}\providecommand{\singleletter}[1]{#1}%
\begin{thebibliography}{30}%
\makeatletter
\providecommand \@ifxundefined [1]{%
 \@ifx{#1\undefined}
}%
\providecommand \@ifnum [1]{%
 \ifnum #1\expandafter \@firstoftwo
 \else \expandafter \@secondoftwo
 \fi
}%
\providecommand \@ifx [1]{%
 \ifx #1\expandafter \@firstoftwo
 \else \expandafter \@secondoftwo
 \fi
}%
\providecommand \natexlab [1]{#1}%
\providecommand \enquote  [1]{``#1''}%
\providecommand \bibnamefont  [1]{#1}%
\providecommand \bibfnamefont [1]{#1}%
\providecommand \citenamefont [1]{#1}%
\providecommand \href@noop [0]{\@secondoftwo}%
\providecommand \href [0]{\begingroup \@sanitize@url \@href}%
\providecommand \@href[1]{\@@startlink{#1}\@@href}%
\providecommand \@@href[1]{\endgroup#1\@@endlink}%
\providecommand \@sanitize@url [0]{\catcode `\\12\catcode `\$12\catcode
  `\&12\catcode `\#12\catcode `\^12\catcode `\_12\catcode `\%12\relax}%
\providecommand \@@startlink[1]{}%
\providecommand \@@endlink[0]{}%
\providecommand \url  [0]{\begingroup\@sanitize@url \@url }%
\providecommand \@url [1]{\endgroup\@href {#1}{\urlprefix }}%
\providecommand \urlprefix  [0]{URL }%
\providecommand \Eprint [0]{\href }%
\providecommand \doibase [0]{http://dx.doi.org/}%
\providecommand \selectlanguage [0]{\@gobble}%
\providecommand \bibinfo  [0]{\@secondoftwo}%
\providecommand \bibfield  [0]{\@secondoftwo}%
\providecommand \translation [1]{[#1]}%
\providecommand \BibitemOpen [0]{}%
\providecommand \bibitemStop [0]{}%
\providecommand \bibitemNoStop [0]{.\EOS\space}%
\providecommand \EOS [0]{\spacefactor3000\relax}%
\providecommand \BibitemShut  [1]{\csname bibitem#1\endcsname}%
\let\auto@bib@innerbib\@empty
\bibitem [{\citenamefont {Kheirandish}, \citenamefont {Javan},\ and\
  \citenamefont {Mohammadzadeh}(2020)}]{kheirandish2020analytical}%
  \BibitemOpen
  \bibfield  {author} {\bibinfo {author} {\bibfnamefont {A.}~\bibnamefont
  {Kheirandish}}, \bibinfo {author} {\bibfnamefont {N.~S.}\ \bibnamefont
  {Javan}}, \ and\ \bibinfo {author} {\bibfnamefont {H.}~\bibnamefont
  {Mohammadzadeh}},\ }\bibfield  {title} {\enquote {\bibinfo {title}
  {Analytical approach to the surface plasmon resonance characteristic of metal
  nanoparticle dimer in dipole-dipole approximation},}\ }\href@noop {}
  {\bibfield  {journal} {\bibinfo  {journal} {Plasmonics}\ }\textbf {\bibinfo
  {volume} {15}},\ \bibinfo {pages} {1807--1814} (\bibinfo {year}
  {2020})}\BibitemShut {NoStop}%
\bibitem [{\citenamefont {Jain}, \citenamefont {Huang},\ and\ \citenamefont
  {El-Sayed}(2007)}]{jain2007universal}%
  \BibitemOpen
  \bibfield  {author} {\bibinfo {author} {\bibfnamefont {P.~K.}\ \bibnamefont
  {Jain}}, \bibinfo {author} {\bibfnamefont {W.}~\bibnamefont {Huang}}, \ and\
  \bibinfo {author} {\bibfnamefont {M.~A.}\ \bibnamefont {El-Sayed}},\
  }\bibfield  {title} {\enquote {\bibinfo {title} {On the universal scaling
  behavior of the distance decay of plasmon coupling in metal nanoparticle
  pairs: a plasmon ruler equation},}\ }\href@noop {} {\bibfield  {journal}
  {\bibinfo  {journal} {Nano Letters}\ }\textbf {\bibinfo {volume} {7}},\
  \bibinfo {pages} {2080--2088} (\bibinfo {year} {2007})}\BibitemShut {NoStop}%
\bibitem [{\citenamefont {Rechberger}\ \emph {et~al.}(2003)\citenamefont
  {Rechberger}, \citenamefont {Hohenau}, \citenamefont {Leitner}, \citenamefont
  {Krenn}, \citenamefont {Lamprecht},\ and\ \citenamefont
  {Aussenegg}}]{rechberger2003optical}%
  \BibitemOpen
  \bibfield  {author} {\bibinfo {author} {\bibfnamefont {W.}~\bibnamefont
  {Rechberger}}, \bibinfo {author} {\bibfnamefont {A.}~\bibnamefont {Hohenau}},
  \bibinfo {author} {\bibfnamefont {A.}~\bibnamefont {Leitner}}, \bibinfo
  {author} {\bibfnamefont {J.}~\bibnamefont {Krenn}}, \bibinfo {author}
  {\bibfnamefont {B.}~\bibnamefont {Lamprecht}}, \ and\ \bibinfo {author}
  {\bibfnamefont {F.}~\bibnamefont {Aussenegg}},\ }\bibfield  {title} {\enquote
  {\bibinfo {title} {Optical properties of two interacting gold
  nanoparticles},}\ }\href@noop {} {\bibfield  {journal} {\bibinfo  {journal}
  {Optics communications}\ }\textbf {\bibinfo {volume} {220}},\ \bibinfo
  {pages} {137--141} (\bibinfo {year} {2003})}\BibitemShut {NoStop}%
\bibitem [{\citenamefont {Atay}, \citenamefont {Song},\ and\ \citenamefont
  {Nurmikko}(2004)}]{atay2004strongly}%
  \BibitemOpen
  \bibfield  {author} {\bibinfo {author} {\bibfnamefont {T.}~\bibnamefont
  {Atay}}, \bibinfo {author} {\bibfnamefont {J.-H.}\ \bibnamefont {Song}}, \
  and\ \bibinfo {author} {\bibfnamefont {A.~V.}\ \bibnamefont {Nurmikko}},\
  }\bibfield  {title} {\enquote {\bibinfo {title} {Strongly interacting plasmon
  nanoparticle pairs: from dipole- dipole interaction to conductively coupled
  regime},}\ }\href@noop {} {\bibfield  {journal} {\bibinfo  {journal} {Nano
  letters}\ }\textbf {\bibinfo {volume} {4}},\ \bibinfo {pages} {1627--1631}
  (\bibinfo {year} {2004})}\BibitemShut {NoStop}%
\bibitem [{\citenamefont {Tabor}\ \emph {et~al.}(2009)\citenamefont {Tabor},
  \citenamefont {Murali}, \citenamefont {Mahmoud},\ and\ \citenamefont
  {El-Sayed}}]{tabor2009use}%
  \BibitemOpen
  \bibfield  {author} {\bibinfo {author} {\bibfnamefont {C.}~\bibnamefont
  {Tabor}}, \bibinfo {author} {\bibfnamefont {R.}~\bibnamefont {Murali}},
  \bibinfo {author} {\bibfnamefont {M.}~\bibnamefont {Mahmoud}}, \ and\
  \bibinfo {author} {\bibfnamefont {M.~A.}\ \bibnamefont {El-Sayed}},\
  }\bibfield  {title} {\enquote {\bibinfo {title} {On the use of plasmonic
  nanoparticle pairs as a plasmon ruler: the dependence of the near-field
  dipole plasmon coupling on nanoparticle size and shape},}\ }\href@noop {}
  {\bibfield  {journal} {\bibinfo  {journal} {The Journal of Physical Chemistry
  A}\ }\textbf {\bibinfo {volume} {113}},\ \bibinfo {pages} {1946--1953}
  (\bibinfo {year} {2009})}\BibitemShut {NoStop}%
\bibitem [{\citenamefont {Su}\ \emph {et~al.}(2003)\citenamefont {Su},
  \citenamefont {Wei}, \citenamefont {Zhang}, \citenamefont {Mock},
  \citenamefont {Smith},\ and\ \citenamefont {Schultz}}]{su2003interparticle}%
  \BibitemOpen
  \bibfield  {author} {\bibinfo {author} {\bibfnamefont {K.-H.}\ \bibnamefont
  {Su}}, \bibinfo {author} {\bibfnamefont {Q.-H.}\ \bibnamefont {Wei}},
  \bibinfo {author} {\bibfnamefont {X.}~\bibnamefont {Zhang}}, \bibinfo
  {author} {\bibfnamefont {J.}~\bibnamefont {Mock}}, \bibinfo {author}
  {\bibfnamefont {D.~R.}\ \bibnamefont {Smith}}, \ and\ \bibinfo {author}
  {\bibfnamefont {S.}~\bibnamefont {Schultz}},\ }\bibfield  {title} {\enquote
  {\bibinfo {title} {Interparticle coupling effects on plasmon resonances of
  nanogold particles},}\ }\href@noop {} {\bibfield  {journal} {\bibinfo
  {journal} {Nano letters}\ }\textbf {\bibinfo {volume} {3}},\ \bibinfo {pages}
  {1087--1090} (\bibinfo {year} {2003})}\BibitemShut {NoStop}%
\bibitem [{\citenamefont {Stauber}(2014)}]{stauber2014plasmonics}%
  \BibitemOpen
  \bibfield  {author} {\bibinfo {author} {\bibfnamefont {T.}~\bibnamefont
  {Stauber}},\ }\bibfield  {title} {\enquote {\bibinfo {title} {Plasmonics in
  dirac systems: from graphene to topological insulators},}\ }\href@noop {}
  {\bibfield  {journal} {\bibinfo  {journal} {Journal of Physics: Condensed
  Matter}\ }\textbf {\bibinfo {volume} {26}},\ \bibinfo {pages} {123201}
  (\bibinfo {year} {2014})}\BibitemShut {NoStop}%
\bibitem [{\citenamefont {Politano}\ \emph {et~al.}(2015)\citenamefont
  {Politano}, \citenamefont {Silkin}, \citenamefont {Nechaev}, \citenamefont
  {Vitiello}, \citenamefont {Viti}, \citenamefont {Aliev}, \citenamefont
  {Babanly}, \citenamefont {Chiarello}, \citenamefont {Echenique},\ and\
  \citenamefont {Chulkov}}]{politano2015interplay}%
  \BibitemOpen
  \bibfield  {author} {\bibinfo {author} {\bibfnamefont {A.}~\bibnamefont
  {Politano}}, \bibinfo {author} {\bibfnamefont {V.}~\bibnamefont {Silkin}},
  \bibinfo {author} {\bibfnamefont {I.}~\bibnamefont {Nechaev}}, \bibinfo
  {author} {\bibfnamefont {M.}~\bibnamefont {Vitiello}}, \bibinfo {author}
  {\bibfnamefont {L.}~\bibnamefont {Viti}}, \bibinfo {author} {\bibfnamefont
  {Z.}~\bibnamefont {Aliev}}, \bibinfo {author} {\bibfnamefont
  {M.}~\bibnamefont {Babanly}}, \bibinfo {author} {\bibfnamefont
  {G.}~\bibnamefont {Chiarello}}, \bibinfo {author} {\bibfnamefont
  {P.}~\bibnamefont {Echenique}}, \ and\ \bibinfo {author} {\bibfnamefont
  {E.}~\bibnamefont {Chulkov}},\ }\bibfield  {title} {\enquote {\bibinfo
  {title} {Interplay of surface and dirac plasmons in topological insulators:
  the case of bi 2 se 3},}\ }\href@noop {} {\bibfield  {journal} {\bibinfo
  {journal} {Physical review letters}\ }\textbf {\bibinfo {volume} {115}},\
  \bibinfo {pages} {216802} (\bibinfo {year} {2015})}\BibitemShut {NoStop}%
\bibitem [{\citenamefont {Politano}, \citenamefont {Lamuta},\ and\
  \citenamefont {Chiarello}(2017)}]{politano2017cutting}%
  \BibitemOpen
  \bibfield  {author} {\bibinfo {author} {\bibfnamefont {A.}~\bibnamefont
  {Politano}}, \bibinfo {author} {\bibfnamefont {C.}~\bibnamefont {Lamuta}}, \
  and\ \bibinfo {author} {\bibfnamefont {G.}~\bibnamefont {Chiarello}},\
  }\bibfield  {title} {\enquote {\bibinfo {title} {Cutting a gordian knot:
  Dispersion of plasmonic modes in bi2se3 topological insulator},}\ }\href@noop
  {} {\bibfield  {journal} {\bibinfo  {journal} {Applied Physics Letters}\
  }\textbf {\bibinfo {volume} {110}},\ \bibinfo {pages} {211601} (\bibinfo
  {year} {2017})}\BibitemShut {NoStop}%
\bibitem [{\citenamefont {Kogar}\ \emph {et~al.}(2015)\citenamefont {Kogar},
  \citenamefont {Vig}, \citenamefont {Thaler}, \citenamefont {Wong},
  \citenamefont {Xiao}, \citenamefont {Reig-i Plessis}, \citenamefont {Cho},
  \citenamefont {Valla}, \citenamefont {Pan}, \citenamefont {Schneeloch} \emph
  {et~al.}}]{kogar2015surface}%
  \BibitemOpen
  \bibfield  {author} {\bibinfo {author} {\bibfnamefont {A.}~\bibnamefont
  {Kogar}}, \bibinfo {author} {\bibfnamefont {S.}~\bibnamefont {Vig}}, \bibinfo
  {author} {\bibfnamefont {A.}~\bibnamefont {Thaler}}, \bibinfo {author}
  {\bibfnamefont {M.}~\bibnamefont {Wong}}, \bibinfo {author} {\bibfnamefont
  {Y.}~\bibnamefont {Xiao}}, \bibinfo {author} {\bibfnamefont {D.}~\bibnamefont
  {Reig-i Plessis}}, \bibinfo {author} {\bibfnamefont {G.}~\bibnamefont {Cho}},
  \bibinfo {author} {\bibfnamefont {T.}~\bibnamefont {Valla}}, \bibinfo
  {author} {\bibfnamefont {Z.}~\bibnamefont {Pan}}, \bibinfo {author}
  {\bibfnamefont {J.}~\bibnamefont {Schneeloch}},  \emph {et~al.},\ }\bibfield
  {title} {\enquote {\bibinfo {title} {Surface collective modes in the
  topological insulators bi 2 se 3 and bi 0.5 sb 1.5 te 3- x se x},}\
  }\href@noop {} {\bibfield  {journal} {\bibinfo  {journal} {Physical review
  letters}\ }\textbf {\bibinfo {volume} {115}},\ \bibinfo {pages} {257402}
  (\bibinfo {year} {2015})}\BibitemShut {NoStop}%
\bibitem [{\citenamefont {Di~Pietro}\ \emph {et~al.}(2013)\citenamefont
  {Di~Pietro}, \citenamefont {Ortolani}, \citenamefont {Limaj}, \citenamefont
  {Di~Gaspare}, \citenamefont {Giliberti}, \citenamefont {Giorgianni},
  \citenamefont {Brahlek}, \citenamefont {Bansal}, \citenamefont {Koirala},
  \citenamefont {Oh} \emph {et~al.}}]{di2013observation}%
  \BibitemOpen
  \bibfield  {author} {\bibinfo {author} {\bibfnamefont {P.}~\bibnamefont
  {Di~Pietro}}, \bibinfo {author} {\bibfnamefont {M.}~\bibnamefont {Ortolani}},
  \bibinfo {author} {\bibfnamefont {O.}~\bibnamefont {Limaj}}, \bibinfo
  {author} {\bibfnamefont {A.}~\bibnamefont {Di~Gaspare}}, \bibinfo {author}
  {\bibfnamefont {V.}~\bibnamefont {Giliberti}}, \bibinfo {author}
  {\bibfnamefont {F.}~\bibnamefont {Giorgianni}}, \bibinfo {author}
  {\bibfnamefont {M.}~\bibnamefont {Brahlek}}, \bibinfo {author} {\bibfnamefont
  {N.}~\bibnamefont {Bansal}}, \bibinfo {author} {\bibfnamefont
  {N.}~\bibnamefont {Koirala}}, \bibinfo {author} {\bibfnamefont
  {S.}~\bibnamefont {Oh}},  \emph {et~al.},\ }\bibfield  {title} {\enquote
  {\bibinfo {title} {Observation of dirac plasmons in a topological
  insulator},}\ }\href@noop {} {\bibfield  {journal} {\bibinfo  {journal}
  {Nature nanotechnology}\ }\textbf {\bibinfo {volume} {8}},\ \bibinfo {pages}
  {556--560} (\bibinfo {year} {2013})}\BibitemShut {NoStop}%
\bibitem [{\citenamefont {Zhang}, \citenamefont {Wang},\ and\ \citenamefont
  {Zhang}(2010)}]{zhang2010topological}%
  \BibitemOpen
  \bibfield  {author} {\bibinfo {author} {\bibfnamefont {X.}~\bibnamefont
  {Zhang}}, \bibinfo {author} {\bibfnamefont {J.}~\bibnamefont {Wang}}, \ and\
  \bibinfo {author} {\bibfnamefont {S.-C.}\ \bibnamefont {Zhang}},\ }\bibfield
  {title} {\enquote {\bibinfo {title} {Topological insulators for
  high-performance terahertz to infrared applications},}\ }\href@noop {}
  {\bibfield  {journal} {\bibinfo  {journal} {Physical review B}\ }\textbf
  {\bibinfo {volume} {82}},\ \bibinfo {pages} {245107} (\bibinfo {year}
  {2010})}\BibitemShut {NoStop}%
\bibitem [{\citenamefont {Autore}\ \emph {et~al.}(2017)\citenamefont {Autore},
  \citenamefont {Di~Pietro}, \citenamefont {Di~Gaspare}, \citenamefont
  {D’Apuzzo}, \citenamefont {Giorgianni}, \citenamefont {Brahlek},
  \citenamefont {Koirala}, \citenamefont {Oh},\ and\ \citenamefont
  {Lupi}}]{autore2017terahertz}%
  \BibitemOpen
  \bibfield  {author} {\bibinfo {author} {\bibfnamefont {M.}~\bibnamefont
  {Autore}}, \bibinfo {author} {\bibfnamefont {P.}~\bibnamefont {Di~Pietro}},
  \bibinfo {author} {\bibfnamefont {A.}~\bibnamefont {Di~Gaspare}}, \bibinfo
  {author} {\bibfnamefont {F.}~\bibnamefont {D’Apuzzo}}, \bibinfo {author}
  {\bibfnamefont {F.}~\bibnamefont {Giorgianni}}, \bibinfo {author}
  {\bibfnamefont {M.}~\bibnamefont {Brahlek}}, \bibinfo {author} {\bibfnamefont
  {N.}~\bibnamefont {Koirala}}, \bibinfo {author} {\bibfnamefont
  {S.}~\bibnamefont {Oh}}, \ and\ \bibinfo {author} {\bibfnamefont
  {S.}~\bibnamefont {Lupi}},\ }\bibfield  {title} {\enquote {\bibinfo {title}
  {Terahertz plasmonic excitations in bi2se3 topological insulator},}\
  }\href@noop {} {\bibfield  {journal} {\bibinfo  {journal} {Journal of
  Physics: Condensed Matter}\ }\textbf {\bibinfo {volume} {29}},\ \bibinfo
  {pages} {183002} (\bibinfo {year} {2017})}\BibitemShut {NoStop}%
\bibitem [{\citenamefont {Politano}, \citenamefont {Viti},\ and\ \citenamefont
  {Vitiello}(2017)}]{politano2017optoelectronic}%
  \BibitemOpen
  \bibfield  {author} {\bibinfo {author} {\bibfnamefont {A.}~\bibnamefont
  {Politano}}, \bibinfo {author} {\bibfnamefont {L.}~\bibnamefont {Viti}}, \
  and\ \bibinfo {author} {\bibfnamefont {M.~S.}\ \bibnamefont {Vitiello}},\
  }\bibfield  {title} {\enquote {\bibinfo {title} {Optoelectronic devices,
  plasmonics, and photonics with topological insulators},}\ }\href@noop {}
  {\bibfield  {journal} {\bibinfo  {journal} {APL Materials}\ }\textbf
  {\bibinfo {volume} {5}},\ \bibinfo {pages} {035504} (\bibinfo {year}
  {2017})}\BibitemShut {NoStop}%
\bibitem [{\citenamefont {Tang}\ \emph {et~al.}(2018)\citenamefont {Tang},
  \citenamefont {Politano}, \citenamefont {Guo}, \citenamefont {Guo},
  \citenamefont {Liu}, \citenamefont {Wang}, \citenamefont {Chen},\ and\
  \citenamefont {Lu}}]{tang2018ultrasensitive}%
  \BibitemOpen
  \bibfield  {author} {\bibinfo {author} {\bibfnamefont {W.}~\bibnamefont
  {Tang}}, \bibinfo {author} {\bibfnamefont {A.}~\bibnamefont {Politano}},
  \bibinfo {author} {\bibfnamefont {C.}~\bibnamefont {Guo}}, \bibinfo {author}
  {\bibfnamefont {W.}~\bibnamefont {Guo}}, \bibinfo {author} {\bibfnamefont
  {C.}~\bibnamefont {Liu}}, \bibinfo {author} {\bibfnamefont {L.}~\bibnamefont
  {Wang}}, \bibinfo {author} {\bibfnamefont {X.}~\bibnamefont {Chen}}, \ and\
  \bibinfo {author} {\bibfnamefont {W.}~\bibnamefont {Lu}},\ }\bibfield
  {title} {\enquote {\bibinfo {title} {Ultrasensitive room-temperature
  terahertz direct detection based on a bismuth selenide topological
  insulator},}\ }\href@noop {} {\bibfield  {journal} {\bibinfo  {journal}
  {Advanced Functional Materials}\ }\textbf {\bibinfo {volume} {28}},\ \bibinfo
  {pages} {1801786} (\bibinfo {year} {2018})}\BibitemShut {NoStop}%
\bibitem [{\citenamefont {Viti}\ \emph {et~al.}(2016)\citenamefont {Viti},
  \citenamefont {Coquillat}, \citenamefont {Politano}, \citenamefont {Kokh},
  \citenamefont {Aliev}, \citenamefont {Babanly}, \citenamefont {Tereshchenko},
  \citenamefont {Knap}, \citenamefont {Chulkov},\ and\ \citenamefont
  {Vitiello}}]{viti2016plasma}%
  \BibitemOpen
  \bibfield  {author} {\bibinfo {author} {\bibfnamefont {L.}~\bibnamefont
  {Viti}}, \bibinfo {author} {\bibfnamefont {D.}~\bibnamefont {Coquillat}},
  \bibinfo {author} {\bibfnamefont {A.}~\bibnamefont {Politano}}, \bibinfo
  {author} {\bibfnamefont {K.~A.}\ \bibnamefont {Kokh}}, \bibinfo {author}
  {\bibfnamefont {Z.~S.}\ \bibnamefont {Aliev}}, \bibinfo {author}
  {\bibfnamefont {M.~B.}\ \bibnamefont {Babanly}}, \bibinfo {author}
  {\bibfnamefont {O.~E.}\ \bibnamefont {Tereshchenko}}, \bibinfo {author}
  {\bibfnamefont {W.}~\bibnamefont {Knap}}, \bibinfo {author} {\bibfnamefont
  {E.~V.}\ \bibnamefont {Chulkov}}, \ and\ \bibinfo {author} {\bibfnamefont
  {M.~S.}\ \bibnamefont {Vitiello}},\ }\bibfield  {title} {\enquote {\bibinfo
  {title} {Plasma-wave terahertz detection mediated by topological insulators
  surface states},}\ }\href@noop {} {\bibfield  {journal} {\bibinfo  {journal}
  {Nano Letters}\ }\textbf {\bibinfo {volume} {16}},\ \bibinfo {pages} {80--87}
  (\bibinfo {year} {2016})}\BibitemShut {NoStop}%
\bibitem [{\citenamefont {Stauber}, \citenamefont {G{\'o}mez-Santos},\ and\
  \citenamefont {Brey}(2013)}]{stauber2013spin}%
  \BibitemOpen
  \bibfield  {author} {\bibinfo {author} {\bibfnamefont {T.}~\bibnamefont
  {Stauber}}, \bibinfo {author} {\bibfnamefont {G.}~\bibnamefont
  {G{\'o}mez-Santos}}, \ and\ \bibinfo {author} {\bibfnamefont
  {L.}~\bibnamefont {Brey}},\ }\bibfield  {title} {\enquote {\bibinfo {title}
  {Spin-charge separation of plasmonic excitations in thin topological
  insulators},}\ }\href@noop {} {\bibfield  {journal} {\bibinfo  {journal}
  {Physical Review B}\ }\textbf {\bibinfo {volume} {88}},\ \bibinfo {pages}
  {205427} (\bibinfo {year} {2013})}\BibitemShut {NoStop}%
\bibitem [{\citenamefont {Autore}\ \emph {et~al.}(2015)\citenamefont {Autore},
  \citenamefont {D'Apuzzo}, \citenamefont {Di~Gaspare}, \citenamefont
  {Giliberti}, \citenamefont {Limaj}, \citenamefont {Roy}, \citenamefont
  {Brahlek}, \citenamefont {Koirala}, \citenamefont {Oh}, \citenamefont
  {Garcia~de Abajo} \emph {et~al.}}]{autore2015plasmon}%
  \BibitemOpen
  \bibfield  {author} {\bibinfo {author} {\bibfnamefont {M.}~\bibnamefont
  {Autore}}, \bibinfo {author} {\bibfnamefont {F.}~\bibnamefont {D'Apuzzo}},
  \bibinfo {author} {\bibfnamefont {A.}~\bibnamefont {Di~Gaspare}}, \bibinfo
  {author} {\bibfnamefont {V.}~\bibnamefont {Giliberti}}, \bibinfo {author}
  {\bibfnamefont {O.}~\bibnamefont {Limaj}}, \bibinfo {author} {\bibfnamefont
  {P.}~\bibnamefont {Roy}}, \bibinfo {author} {\bibfnamefont {M.}~\bibnamefont
  {Brahlek}}, \bibinfo {author} {\bibfnamefont {N.}~\bibnamefont {Koirala}},
  \bibinfo {author} {\bibfnamefont {S.}~\bibnamefont {Oh}}, \bibinfo {author}
  {\bibfnamefont {F.~J.}\ \bibnamefont {Garcia~de Abajo}},  \emph {et~al.},\
  }\bibfield  {title} {\enquote {\bibinfo {title} {Plasmon--phonon interactions
  in topological insulator microrings},}\ }\href@noop {} {\bibfield  {journal}
  {\bibinfo  {journal} {Advanced Optical Materials}\ }\textbf {\bibinfo
  {volume} {3}},\ \bibinfo {pages} {1257--1263} (\bibinfo {year}
  {2015})}\BibitemShut {NoStop}%
\bibitem [{\citenamefont {Venuthurumilli}\ \emph {et~al.}(2019)\citenamefont
  {Venuthurumilli}, \citenamefont {Wen}, \citenamefont {Iyer}, \citenamefont
  {Chen},\ and\ \citenamefont {Xu}}]{venuthurumilli2019near}%
  \BibitemOpen
  \bibfield  {author} {\bibinfo {author} {\bibfnamefont {P.~K.}\ \bibnamefont
  {Venuthurumilli}}, \bibinfo {author} {\bibfnamefont {X.}~\bibnamefont {Wen}},
  \bibinfo {author} {\bibfnamefont {V.}~\bibnamefont {Iyer}}, \bibinfo {author}
  {\bibfnamefont {Y.~P.}\ \bibnamefont {Chen}}, \ and\ \bibinfo {author}
  {\bibfnamefont {X.}~\bibnamefont {Xu}},\ }\bibfield  {title} {\enquote
  {\bibinfo {title} {Near-field imaging of surface plasmons from the bulk and
  surface state of topological insulator bi2te2se},}\ }\href@noop {} {\bibfield
   {journal} {\bibinfo  {journal} {ACS Photonics}\ }\textbf {\bibinfo {volume}
  {6}},\ \bibinfo {pages} {2492--2498} (\bibinfo {year} {2019})}\BibitemShut
  {NoStop}%
\bibitem [{\citenamefont {Wang}\ and\ \citenamefont
  {Law}(2020)}]{wang2020propagating}%
  \BibitemOpen
  \bibfield  {author} {\bibinfo {author} {\bibfnamefont {Y.}~\bibnamefont
  {Wang}}\ and\ \bibinfo {author} {\bibfnamefont {S.}~\bibnamefont {Law}},\
  }\bibfield  {title} {\enquote {\bibinfo {title} {Propagating dirac plasmon
  polaritons in topological insulators},}\ }\href@noop {} {\bibfield  {journal}
  {\bibinfo  {journal} {Journal of Optics}\ }\textbf {\bibinfo {volume} {22}},\
  \bibinfo {pages} {125001} (\bibinfo {year} {2020})}\BibitemShut {NoStop}%
\bibitem [{\citenamefont {Ginley}\ and\ \citenamefont
  {Law}(2016)}]{Ginley2016}%
  \BibitemOpen
  \bibfield  {author} {\bibinfo {author} {\bibfnamefont {T.}~\bibnamefont
  {Ginley}}\ and\ \bibinfo {author} {\bibfnamefont {S.}~\bibnamefont {Law}},\
  }\bibfield  {title} {\enquote {\bibinfo {title} {Growth of bi2se3 topological
  insulator films using a selenium cracker source},}\ }\href {\doibase
  10.1116/1.4941134} {\bibfield  {journal} {\bibinfo  {journal} {Journal of
  Vacuum Science and Technology B}\ }\textbf {\bibinfo {volume} {34}},\
  \bibinfo {pages} {02L105} (\bibinfo {year} {2016})}\BibitemShut {NoStop}%
\bibitem [{\citenamefont {Wang}\ and\ \citenamefont {Law}()}]{daniel}%
  \BibitemOpen
  \bibfield  {author} {\bibinfo {author} {\bibfnamefont {Z.}~\bibnamefont
  {Wang}}\ and\ \bibinfo {author} {\bibfnamefont {S.}~\bibnamefont {Law}},\
  }\bibfield  {title} {\enquote {\bibinfo {title} {Optimization of the growth
  of the van der waals materials bi$_2$se$_3$ and (bi$_0.5$in$_0.5$)$_2$se$_3$
  by molecular beam epitaxy},}\ }\href@noop {} {\bibinfo  {journal}
  {https://arxiv.org/abs/2107.09771}\ }\BibitemShut {NoStop}%
\bibitem [{\citenamefont {Maier}(2007)}]{maier2007plasmonics}%
  \BibitemOpen
\bibfield  {journal} {  }\bibfield  {author} {\bibinfo {author} {\bibfnamefont
  {S.~A.}\ \bibnamefont {Maier}},\ }\href@noop {} {\emph {\bibinfo {title}
  {Plasmonics: fundamentals and applications}}}\ (\bibinfo  {publisher}
  {Springer Science \& Business Media},\ \bibinfo {year} {2007})\BibitemShut
  {NoStop}%
\bibitem [{\citenamefont {Ginley}\ and\ \citenamefont
  {Law}(2018)}]{ginley2018coupled}%
  \BibitemOpen
  \bibfield  {author} {\bibinfo {author} {\bibfnamefont {T.~P.}\ \bibnamefont
  {Ginley}}\ and\ \bibinfo {author} {\bibfnamefont {S.}~\bibnamefont {Law}},\
  }\bibfield  {title} {\enquote {\bibinfo {title} {Coupled dirac plasmons in
  topological insulators},}\ }\href@noop {} {\bibfield  {journal} {\bibinfo
  {journal} {Advanced Optical Materials}\ }\textbf {\bibinfo {volume} {6}},\
  \bibinfo {pages} {1800113} (\bibinfo {year} {2018})}\BibitemShut {NoStop}%
\bibitem [{\citenamefont {Giannini}\ \emph {et~al.}(2011)\citenamefont
  {Giannini}, \citenamefont {Fern{\'a}ndez-Dom{\'\i}nguez}, \citenamefont
  {Heck},\ and\ \citenamefont {Maier}}]{giannini2011plasmonic}%
  \BibitemOpen
  \bibfield  {author} {\bibinfo {author} {\bibfnamefont {V.}~\bibnamefont
  {Giannini}}, \bibinfo {author} {\bibfnamefont {A.~I.}\ \bibnamefont
  {Fern{\'a}ndez-Dom{\'\i}nguez}}, \bibinfo {author} {\bibfnamefont {S.~C.}\
  \bibnamefont {Heck}}, \ and\ \bibinfo {author} {\bibfnamefont {S.~A.}\
  \bibnamefont {Maier}},\ }\bibfield  {title} {\enquote {\bibinfo {title}
  {Plasmonic nanoantennas: fundamentals and their use in controlling the
  radiative properties of nanoemitters},}\ }\href@noop {} {\bibfield  {journal}
  {\bibinfo  {journal} {Chemical reviews}\ }\textbf {\bibinfo {volume} {111}},\
  \bibinfo {pages} {3888--3912} (\bibinfo {year} {2011})}\BibitemShut {NoStop}%
\bibitem [{\citenamefont {Hao}\ \emph {et~al.}(2009)\citenamefont {Hao},
  \citenamefont {Nordlander}, \citenamefont {Sonnefraud}, \citenamefont
  {Dorpe},\ and\ \citenamefont {Maier}}]{hao2009tunability}%
  \BibitemOpen
  \bibfield  {author} {\bibinfo {author} {\bibfnamefont {F.}~\bibnamefont
  {Hao}}, \bibinfo {author} {\bibfnamefont {P.}~\bibnamefont {Nordlander}},
  \bibinfo {author} {\bibfnamefont {Y.}~\bibnamefont {Sonnefraud}}, \bibinfo
  {author} {\bibfnamefont {P.~V.}\ \bibnamefont {Dorpe}}, \ and\ \bibinfo
  {author} {\bibfnamefont {S.~A.}\ \bibnamefont {Maier}},\ }\bibfield  {title}
  {\enquote {\bibinfo {title} {Tunability of subradiant dipolar and fano-type
  plasmon resonances in metallic ring/disk cavities: implications for nanoscale
  optical sensing},}\ }\href@noop {} {\bibfield  {journal} {\bibinfo  {journal}
  {ACS nano}\ }\textbf {\bibinfo {volume} {3}},\ \bibinfo {pages} {643--652}
  (\bibinfo {year} {2009})}\BibitemShut {NoStop}%
\bibitem [{\citenamefont {Wang}\ \emph {et~al.}(2020)\citenamefont {Wang},
  \citenamefont {Ginley}, \citenamefont {Mambakkam}, \citenamefont {Chandan},
  \citenamefont {Zhang}, \citenamefont {Ni},\ and\ \citenamefont
  {Law}}]{wang2020plasmon}%
  \BibitemOpen
  \bibfield  {author} {\bibinfo {author} {\bibfnamefont {Z.}~\bibnamefont
  {Wang}}, \bibinfo {author} {\bibfnamefont {T.~P.}\ \bibnamefont {Ginley}},
  \bibinfo {author} {\bibfnamefont {S.~V.}\ \bibnamefont {Mambakkam}}, \bibinfo
  {author} {\bibfnamefont {G.}~\bibnamefont {Chandan}}, \bibinfo {author}
  {\bibfnamefont {Y.}~\bibnamefont {Zhang}}, \bibinfo {author} {\bibfnamefont
  {C.}~\bibnamefont {Ni}}, \ and\ \bibinfo {author} {\bibfnamefont
  {S.}~\bibnamefont {Law}},\ }\bibfield  {title} {\enquote {\bibinfo {title}
  {Plasmon coupling in topological insulator multilayers},}\ }\href@noop {}
  {\bibfield  {journal} {\bibinfo  {journal} {Physical Review Materials}\
  }\textbf {\bibinfo {volume} {4}},\ \bibinfo {pages} {115202} (\bibinfo {year}
  {2020})}\BibitemShut {NoStop}%
\bibitem [{\citenamefont {Yan}\ \emph {et~al.}(2012)\citenamefont {Yan},
  \citenamefont {Xia}, \citenamefont {Li},\ and\ \citenamefont
  {Avouris}}]{Yan2012}%
  \BibitemOpen
  \bibfield  {author} {\bibinfo {author} {\bibfnamefont {H.}~\bibnamefont
  {Yan}}, \bibinfo {author} {\bibfnamefont {F.}~\bibnamefont {Xia}}, \bibinfo
  {author} {\bibfnamefont {Z.}~\bibnamefont {Li}}, \ and\ \bibinfo {author}
  {\bibfnamefont {P.}~\bibnamefont {Avouris}},\ }\bibfield  {title} {\enquote
  {\bibinfo {title} {{Plasmonics of coupled graphene micro-structures}},}\
  }\href {\doibase 10.1088/1367-2630/14/12/125001} {\bibfield  {journal}
  {\bibinfo  {journal} {New Journal of Physics}\ }\textbf {\bibinfo {volume}
  {14}},\ \bibinfo {pages} {125001} (\bibinfo {year} {2012})},\ \Eprint
  {http://arxiv.org/abs/1205.6841} {1205.6841} \BibitemShut {NoStop}%
\bibitem [{\citenamefont {Win}(2018)}]{exponential}%
  \BibitemOpen
  \bibfield  {author} {\bibinfo {author} {\bibfnamefont {C.~C.}\ \bibnamefont
  {Win}},\ }\bibfield  {title} {\enquote {\bibinfo {title} {Symmetrical
  double-well potential and its application},}\ }\href@noop {} {\  (\bibinfo
  {year} {2018})}\BibitemShut {NoStop}%
\bibitem [{\citenamefont {Harrison}\ and\ \citenamefont
  {Alex}(2016)}]{harrisonquantum}%
  \BibitemOpen
  \bibfield  {author} {\bibinfo {author} {\bibfnamefont {P.}~\bibnamefont
  {Harrison}}\ and\ \bibinfo {author} {\bibfnamefont {V.}~\bibnamefont
  {Alex}},\ }\href@noop {} {\emph {\bibinfo {title} {Quantum Wells, Wires and
  Dots: Theoretical and Computational Physics of Semiconductor
  Nanostructures}}}\ (\bibinfo  {publisher} {Wiley},\ \bibinfo {year}
  {2016})\BibitemShut {NoStop}%
\end{thebibliography}%
\end{document}